\documentclass[11pt]{article}
\usepackage{cite}
\usepackage{amsmath,amsfonts,amssymb}%,calrsfs}
\usepackage[small,bf,hang]{caption}
\usepackage{slashed}
\usepackage{color}

\usepackage{extarrows}
\usepackage{hyperref}

\usepackage{geometry}
\def\hybrid{
        \topmargin -20pt
        \oddsidemargin 0pt
        \headheight 0pt \headsep 0pt
        \textwidth 6.25in % A4 paper
        \textheight 9.5in % A4 paper
        \marginparwidth .875in
        \parskip 5pt plus 1pt \jot = 1.5ex}

\hybrid

 \csname
@addtoreset\endcsname{equation}{section}

\def\moth{\mathsurround=0pt}
\newdimen\zo \zo=0pt

\def\tick{\leaders\hrule height 0.5ex depth 0pt \hskip 0.5pt}
\def\upboxfill{$\moth \setbox\zo\hbox{\tick}%
  \hskip 3pt\hbox to 0pt{$\tick$\hss}\hrulefill \hbox to 7.5pt{$\tick$\hss}$}

\def\dtick{\leaders\hrule height .34pt depth 0.5ex \hskip 0.5pt}
\def\downboxfill{$\moth \setbox\zo\hbox{\dtick}%
  \hskip 2pt\hbox to 0pt{$\dtick$\hss}\hrulefill \hbox to 2pt{$\dtick$\hss}$}

\def\bec{\begin{center}}
\def\ec{\end{center}}

\def\nn{\nonumber}

\def\be{\begin{equation}}
\def\ee{\end{equation}}
\def\bea{\begin{eqnarray}}
\def\eea{\end{eqnarray}}
\def\ba{\begin{array}}
\def\ea{\end{array}}

\begin{document}

\begin{titlepage}
\rightline{}
%\rightline\today
\begin{center}
\vskip 1.5cm
 {\Large \bf{   
 The $\beta$-symmetry of supergravity}}
\vskip 1.7cm

{\large\bf {Walter H. Baron$^*$, Diego Marqu\'es$^\dag$ and Carmen A. Nu\~nez$^\dag$}}
\vskip 1cm

$^*$ {\it  Instituto de F\'isica La Plata (CONICET-UNLP)\\
Departamento de Matem\'atica, Universidad Nacional de La Plata, Argentina. }\\
 
\vskip .3cm

$^\dag$ {\it   Instituto de Astronom\'ia y F\'isica del Espacio (CONICET-UBA), \\
 Departamento de F\'isica, Universidad de Buenos Aires, Argentina.}\\
\vskip .1cm

\vskip .4cm

wbaron@fisica.unlp.edu.ar, diegomarques@iafe.uba.ar, carmen@iafe.uba.ar

\vskip .4cm

\end{center}

\bigskip\bigskip
\begin{center} 
\textbf{Abstract}

\end{center} 
\begin{quote}

Continuous   O$(d,d)$  global  symmetries emerge in 
 Kaluza-Klein reductions of $D$-dimensional  string supergravities to $D-d$ dimensions. We show that the non-geometric elements of this group effectively act in the $D$-dimensional parent theory as a hidden bosonic symmetry that fixes its couplings: the $\beta$-symmetry. We give the explicit $\beta$-transformations to first order in $\alpha'$ and verify the invariance of the action as well as the closure of the transformation rules.

\end{quote} 
\vfill
\setcounter{footnote}{0}
\end{titlepage}

\section{Introduction}

Compactifications of the string effective field theories on $d$-dimensional tori posses a continuous O$(d,d)$ rigid  symmetry  \cite{Maharana} to all orders in $\alpha'$ \cite{Sen}. This symmetry is the footprint of T-duality in the supergravity limit. 

 The couplings in the higher derivative expansion of the string (super)gravities can then be predicted  by
demanding the emergence of O$(d,d)$ symmetries after compactification. Although this procedure is in general tedious, as it requires non-trivial field redefinitions to make the symmetry manifest, it  has been successfully pursued up to order $\alpha'{}^3$ \cite{Garousi}.  
An alternative procedure explores symmetry principles that determine double field theory interactions, either through higher derivative deformations of generalized diffeomorphisms \cite{HSZ} or double-Lorentz symmetries \cite{MN,Baron:2018lve} . The invariant action can then be  downgraded to supergravity with all the couplings fixed. 
 
The former method involves heavy brute force  computations that become non-viable after a few orders, while the latter is currently confronted with an obstruction  starting at the quartic Riemann interactions common to all string theories \cite{HW}\footnote{The claim is that the obstruction applies to the background independent frame-like formulation of double field theory with the strong constraint  \cite{Siegel, HK}. Double field theory on tori with the weak constraint \cite{Hull} is not expected to face any obstructions,  since it is a perturbative expansion of string field theory.}. We are then at a stage that requires simplifications in the first approach, and clarifications in the second one.
 
The key observation introduced in this paper is that the appearance of O$(d,d)$ symmetries in the $D-d$ dimensional theory can be assessed already in the $D$ dimensional parent action. The idea is extremely simple and goes as follows. Starting with a  string effective field theory in $D$ dimensions, the  Kaluza-Klein reduction to $D-d$ dimensions, keeping only the massless modes, consists of three steps: 
\begin{itemize}
    \item Split the $D$ space-time coordinates into $D-d$ external and $d$ internal directions, and impose that the fields are independent of the internal ones.
    
    \item Propose a Kaluza-Klein parametrization of the higher dimensional fields in terms of those in lower dimensions. The purpose of this step is to obtain fields with standard transformation properties with respect to the local symmetries.
    
    \item Enforce higher-derivative field redefinitions that allow assembling the degrees of freedom into O$(d,d)$ multiplets, so as to make the O$(d,d)$  symmetry manifest and not corrected by higher derivatives. In some cases this requires including extra gauge degrees of freedom \cite{Gaugedof}.
\end{itemize}

The last two items are just field redefinitions. What they do is to take the $D-d$ effective action  obtained directly from the $D$ dimensional one, in which derivatives are non-vanishing only in the external directions, to a scheme in which the symmetries are manifest. These redefinitions are purely aesthetical, since the symmetries,  though hidden, are still there.  Hence, there must be a way to identify the O$(d,d)$ symmetry  directly in the $D$ dimensional action.  This is what we will show in this paper.

While the {\it geometric} subgroup  of O$(d,d)$, consisting of rigid $d$-dimensional diffeomorphisms and two-form shifts, acts trivially with no higher-derivative corrections, the {\it non-geometric} sector parametrized by a bi-vector $\beta$ fixes entirely the effective action in the scheme in which  it looks exactly like the higher-dimensional theory. In other words, the non-geometric sector fixes the higher-dimensional action, and it does so by acting effectively as if it were a symmetry in $D$ dimensions. 

The paper is organized as follows. In Section 2 we expose the $\beta$-invariance of the two-derivative universal string supergravity. In section 3 we derive the first order $\alpha'$ corrections to the $\beta$-transformations in the generalized Bergshoeff-de Roo scheme and verify closure together with the local symmetries.

\section{The $\beta$-symmetry to lowest order}

Each term in the universal  two derivative NS-NS action 
\be
S = \int d^{D}x \sqrt{-g} e^{-2\phi} \left(R - 4 (\nabla \phi)^2  + 4 \Box \phi - \frac 1 {12} H^2\right) \  \label{DSugra}
\ee
is manifestly invariant under local $D$ dimensional diffeomorphisms and gauge transformations of the   two-form. These symmetries in turn contain GL$(D) \times {\rm R}^{\frac {D(D-1)} 2}$ as a rigid continuous subgroup, infinitesimally parametrized by $a^\mu{}_\nu$ and $B_{\mu \nu}$ acting on $E_{\mu \nu} = g_{\mu \nu} + b_{\mu \nu}$ and $\phi$ as follows
\begin{subequations}
\begin{align}
\delta \partial_{\mu} &= - a^\rho{}_\mu \partial_\rho \, ,   \\
\delta E_{\mu \nu} &= B_{\mu \nu} - a^{\rho}{}_{\mu} E_{\rho \nu} - a^{\rho}{}_\nu E_{\mu \rho}  \, ,  \\ 
\delta \phi &= -\frac 1 2 a^\mu{}_\mu  \, . 
\end{align}
\end{subequations}
This is the geometric subgroup of O$(D,D)$, which additionally contains non-geometric elements parametrized by  a constant bi-vector $\beta^{\mu \nu}$
\begin{subequations} \label{BetaSymmetry}
\begin{align}
\delta E_{\mu \nu} &= - E_{\mu \rho} \beta^{\rho \sigma} E_{\sigma \nu} \, ,\\ 
\delta \phi &= \frac 1 2 \beta^{\mu \nu} E_{\mu \nu}\, .
\end{align}
\end{subequations}
These non-geometric transformations are {\it not}  symmetries of supergravity (\ref{DSugra}). Demanding invariance under the full O$(D,D)$ group requires doubling the space-time coordinates
and  adding extra terms in the action, as is known from double field theory \cite{Hohm:2010jy}. This is not the route that we follow in this paper: here we deal  with pure  supergravity.  

Even if $D$ dimensional supergravity is not invariant under  O$(D,D)$, we know that its compactification on $T^d$ must be O$(d,d) \in {\rm O}(D,D)$ symmetric. Operationally the compactification amounts to the  assumption that the fields do not depend on the internal directions, which implies truncating the {\it derivatives} to be purely {\it external}. In such case, the action gains the full O$(d,d)$ symmetry, given by the trivial embedding into O$(D,D)$ such that the {\it parameters} contain only {\it internal} components. Then,  (\ref{BetaSymmetry}) effectively becomes a symmetry of (\ref{DSugra}) under the constraint
\be
\beta^{\mu \nu} \partial_\nu \dots = 0 \ . \label{BetaConstraint}
\ee
As a consequence, the O$(d,d)$ symmetry of (\ref{DSugra}) compactified on $T^d$ can be determined, for all practical purposes, directly in (\ref{DSugra}) through the action of (\ref{BetaSymmetry}) constrained as in (\ref{BetaConstraint}).

Checking the $\beta$-invariance of the action turns out to be easier in the frame formulation, where flattening the indices of the fields with the frame, and defining flattened variations 
\be
\delta e_{a b} = e^\mu{}_a \delta e_{\mu b} \ ,  \ \ \ \ \delta b_{a b} = e^\mu{}_a e^\nu{}_b \delta b_{\mu \nu} \ ,
\ee
the transformations take the form
\be
\delta e_{a b}= - b_{a c} \beta^c{}_b \ , \ \ \ 
\delta b_{a b}= - \beta_{a b} - b_{a c} \beta^{c d} b_{d b}  \ , \ \ \ 
\delta \phi =  \frac 1 2 \delta e_a{}^a  \ . \label{betatransf}
\ee
These in turn dictate the variations of the tensors and connections that appear in the action (see the Appendix for details on the notation)
\begin{subequations}
\begin{align}
[\delta, D_a] &= 0 \, , \\
\delta w_{c a b} &= \beta_{[a}{}^dH_{b]cd}-\frac12\beta_c{}^dH_{abd} \, , \\
\delta  H_{abc}&= 6w^d{}_{[ac}\beta_{b]d}\, ,  \\
\delta (\nabla_a\phi)&=\frac12\beta^{cd}H_{acd} \, , \\
\delta ( \nabla_a \nabla_b\phi)&=\frac12\beta^{cd}\nabla_{(a}H_{b)cd}-\beta^{ce}w_{e(a}{}^dH_{b)cd} - \beta^c{}_{(a} H_{b) c d} \nabla^d \phi \, .
\end{align}
\end{subequations}
To derive these expressions we have used (\ref{BetaConstraint}) and the fact that $\beta^{\mu \nu}$ is constant and antisymmetric, which in turn imply
\be
 D_a \beta^{ b c} = 4 \beta^{d [b} \omega_{[d a]}{}^{c]} \ , \ \ \  \qquad \beta^{a b} \omega_{a b c} = 0 \ .
\ee

To prove the invariance of the action (\ref{DSugra}) is now trivial, taking into account that the above transformations yield
\begin{subequations}
\begin{align} \delta \left( \sqrt{- g} e^{- 2 \phi} \right) &= 0 \, , \\
\delta  R &= -2 \beta^{cd} \nabla^b H_{bcd} + 5 \beta^{c d} \omega_{cab}H_d{}^{ab} \, , \\
\delta (\nabla \phi)^2 &= \beta^{c d} H_{b c d} \nabla^b \phi \, ,  \\
\delta \Box \phi &= \frac 1 2 \beta^{cd} \nabla^b H_{bcd} - \beta^{c d} \omega_{cab}H_d{}^{ab} +\beta^{c d} H_{b c d} \nabla^b \phi \, ,  \\
\delta H^2  &=12\, \beta^{c d} \omega_{c a b} H_d{}^{a b}\, .
\end{align}
\end{subequations}
In fact, the $\beta$-invariance of a generic combination of terms preserved by the local symmetries
\begin{eqnarray}
0\ &=& \delta \left(R + m\, (\nabla \phi)^2  + n\, \Box \phi + p\, H^2 \right) \\
&=& \beta^{c d}  \nabla^b H_{b c d} \left(-2 + \frac n 2\right) + \beta^{ c d} \omega_{c a b} H_{d}{}^{a b} \left(5 - n + 12 p \right)   + \ \beta^{c d} H_{b c d} \nabla^b \phi \left(m + n\right)\ ,\nn
\end{eqnarray}
fixes the value of the coefficients to 
\be
m = -4 \ , \ \ \ n = 4 \ , \ \ \ p = - \frac 1 {12} \ , 
\ee
selecting (\ref{DSugra}) as the unique $\beta$-symmetric theory. 

Together with Lorentz transformations, \eqref{betatransf}  close into the bracket
\be
\left[\delta_1\, , \, \delta_2\right] = - \delta_{12} \ , \ \ \ \Lambda_{1 2 a b} = 2 \, \beta_{1[a}{}^{c} \beta_{2\, b] c} 
    + 2\, \Lambda_{1 [a}{}^{c} \Lambda_{2\, b] c} \ .
\ee

\section{The $\beta$-symmetry to first order}

 In the bi-parametric $(a\, ,\, b)$ generalized Bergshoeff-de Roo scheme, all  string effective actions up to first order in $\alpha'$ are included in \cite{MN}
\begin{equation}
    S= \int d^D x \sqrt{-g} e^{-2\phi} \left(L^{(0)} + a L^{(1)}_a + b L^{(1)}_b \right) \ , \label{bdrac}
\end{equation}
where the lowest order Lagrangian $L^{(0)}$ is defined in  (\ref{DSugra}), and the first order one can be written in a flattened fashion with
\begin{subequations}
\begin{align}
L^{(1)}_a &= \frac 1 4 H^{a b c} \Omega^{(-)}_{a b c} - \frac 1 8 R^{(-)}_{a b c d} R^{(-) a b c d} \ , \\
L^{(1)}_b &= - \frac 1 4 H^{a b c} \Omega^{(+)}_{a b c} - \frac 1 8 R^{(+)}_{a b c d} R^{(+) a b c d} \ .
\end{align}
\end{subequations}
Defining $\omega^{(\pm)}_{a b c} = \omega_{a b c} \pm \frac 1 2 H_{a b c}$, these expressions contain
\begin{eqnarray}
\Omega^{(\pm)}_{a b c} &=&  \omega^{(\pm)}_{[\underline{a} d}{}^e D_{\underline{b}}\omega^{(\pm)}_{\underline{c}] e}{}^d + \omega^{(\pm)}_{[\underline{a} d}{}^e\omega^{(\pm)}_{f e}{}^d\omega_{\underline{b c}]}{}^f + \frac 2 3 \omega^{(\pm)}_{[\underline{a} d}{}^e\omega^{(\pm)}_{\underline{b} e}{}^f\omega^{(\pm)}_{\underline{c}] f}{}^d \ ,\\
R^{(\pm)}_{a b c d} &=& 2 D_{[a}\omega^{(\pm)}_{b]cd} + 2 \omega_{[a b] }{}^e \omega^{(\pm)}_{e c d}  + 2 \omega^{(\pm)}_{[\underline{a} c}{}^e \omega^{(\pm)}_{\underline{b}]e d} \ .
\end{eqnarray}

We look for measure preserving $\beta$-transformations that tie the variation of the dilaton to that of the frame field to all orders
\begin{equation}
\delta  \left(\sqrt{-g} e^{-2\phi}\right) = 0 \ \ \ \ \Rightarrow \ \ \ \ \delta \phi =  \frac 1 2 \delta e_a{}^a \ .
\end{equation}
The $\beta$-invariance up to first order is then guaranteed by
\begin{equation}
\int d^D x \sqrt{-g} e^{-2\phi} \left(\delta^{(1)} L^{(0)} + a \, \delta^{(0)}  L^{(1)}_a + b\, \delta^{(0)} L^{(1)}_b \right) = 0 \ , \label{firstorderinvariance}
\end{equation}
where $\delta^{(0)}$ denotes the lowest order variations of the previous section.

To find $\delta^{(1)}$, the first order  $\alpha'$-corrections to the $\beta$-transformations, we will consider an expansion in powers of the fluxes $\omega_{a b c}$, $H_{a b c}$ and $D_{a}{\phi}$. This is a useful strategy that serves as an organizing principle, mimicking a background field expansion. The difference is that fluxes are composite fields, and hence obey Bianchi identities (BI) that relate different orders, namely  (\ref{BI1}), (\ref{BI2}) and (\ref{BI3}). To remove ambiguities, one uses the leading terms in the BI to take the leading order to a minimal form at the expense of introducing subleading terms. Once the leading order is fixed, one moves to the next order and again takes it to a minimal form using BI at the expense of inducing further higher order terms, and so on. As an example, the lowest order equations of motion admit a flux expansion of the form
\begin{subequations} \label{lowesteom}
\begin{align}
\Delta b_{a b} &= \frac 1 2 \nabla_c H^c{}_{a b} - \nabla_c \phi H^c{}_{a b} = \frac 1 2 D_c H^c{}_{a b} + \dots \label{lowestbeom}\\
\Delta e_{a b} &= - 2 \left(R_{a b} + 2 \nabla_{(a} \nabla_{b)} \phi - \frac 1 4 H_{a c d} H_b{}^{c d}\right) = - 4 D_{a}D_b\phi - 2 D_a \omega_{c b}{}^c + 2 D_c \omega_{a b}{}^c + \dots \, ,\label{lowesteeom}
\end{align}
\end{subequations}
where the dots represent quadratic terms, which are subleading with respect to those that we have written explicitly. The way the lowest order in (\ref{lowesteeom}) looks like can be changed using the BI (\ref{BI3}), but once it is fixed, the subleading terms are also fixed. Note that flat derivatives commute at leading order. 

 Integrating by parts, the first term in (\ref{firstorderinvariance}) can be taken to the form
\begin{eqnarray}
\int d^D x \sqrt{-g} e^{-2\phi} \delta^{(1)} L^{(0)} &=& \int d^D x \sqrt{-g} e^{-2\phi} \left( \delta^{(1)}b^{a b} \Delta b_{a b} + \delta^{(1)} e^{a b} \Delta e_{a b}
\right) \, .\label{d1L0}
\end{eqnarray}
On the other hand, since  the lowest order transformation rules (\ref{betatransf}) are  known, we can readily compute $\delta^{(0)} L^{(1)}$ and determine the first order deformations by requiring invariance of the action (\ref{firstorderinvariance}).
To this end, it is convenient  to consider the particular case $b = 0$, and then infer the general transformations from the fact that $L_b^{(1)} = L_a^{(1)} [H \to -H]$. The leading terms in the flux expansion of $\delta^{(0)} L_a^{(1)}$ turn out to be cubic, i.e.
\begin{eqnarray}
\delta^{(0)}  L^{(1)}_a = \sum_{h = 0}^3 \left[\delta^{(0)}  L^{(1)}_a\right]{}_{(h, 3-h)} + \dots \ ,
\end{eqnarray}
where $(h, 3-h)$ denotes terms with $h$ fluxes  $H$ and $3-h$  fluxes $\omega$, and the dots represent subleading expressions. Each term in this expansion can be taken to the form
\begin{subequations}\label{d0L1a}
\begin{align}
\left[\delta^{(0)}  L^{(1)}_a\right]{}_{(0, 3)} &= \Delta e^{a b} \left[ \frac 1 4  \beta_a{}^c  \omega_{b d e} \omega_c{}^{de} \right] + [{\cal D}_{a} T^a]_{(0,3)}
\, ,\\
\left[\delta^{(0)}  L^{(1)}_a\right]{}_{(1, 2)} &= \Delta b^{a b} \left[ -\beta^{e c} \omega_{ea}{}^d \omega_{b c d} + \beta^{e c} \omega_{a e}{}^d \omega_{b c d} + \frac 1 2 \beta_a{}^c \omega_{b d e} \omega_c{}^{d e}\right]\nonumber \\
& \ \ \ + \Delta e^{a b} \left[- \frac 1 8 \beta_a{}^e \omega_{b c d} H_e{}^{c d} - \frac 1 8 \beta_a{}^e H_{b c d} \omega_e{}^{c d}\right] + [{\cal  D}_a T^a]_{(1,2)} \, , \\
\left[\delta^{(0)}  L^{(1)}_a\right]{}_{(2, 1)} &= \Delta b^{a b} \left[ \frac 1 2 \beta^{e c} \omega_{e a}{}^d H_{b c d} - \frac 1 2 \beta^{e c} \omega_{a e}{}^d H_{b c d} - \frac 1 2 \beta_a{}^c \omega_{b d e} H_c{}^{d e} - \frac 1 2 \beta_a{}^c H_{b d e} \omega_{c}{}^{d e}\right] \nonumber\\ & \ \ \ + \Delta e^{a b} \left[\frac 1 {16} \beta_a{}^e H_{b c d} H_e{}^{c d}\right] + [{\cal D}_a T^a]_{(2,1)} \, ,\\
\left[\delta^{(0)}  L^{(1)}_a\right]{}_{(3, 0)} &= \Delta b^{a b} \left[ \frac 1 8 \beta_a{}^c H_{b d e} H_c{}^{de}\right]  + [{\cal  D}_{a}T^a]_{(3,0)} \, ,
\end{align}
\end{subequations}
where $\Delta b_{a b}$ and $\Delta e_{a b}$ contain the leading order of the equations of motion (\ref{lowesteom}). The derivative ${\cal D}_{a}T^a$   gives rise to a total derivative when introduced in the action
\be
{\cal D}_a T^{a} = D_a T^a - 2 D_a \phi T^a -\omega_{b a }{}^b T^a \ , \ \ \ \ \sqrt{-g} e^{-2 \phi} {\cal D}_a T^a = \partial_\mu \left(\sqrt{-g} e^{-2 \phi} e^\mu{}_a T^a\right) \ , 
 \ee
and hence it is not relevant for our purposes. Nevertheless,  for completeness  we give the explicit expression of the vector $T^a$ to cubic order in  \eqref{ta}.

Written like this, it is now trivial to extract the first order corrections to the $\beta$-transformations proportional to the parameter $a$,  introducing (\ref{d1L0}) and (\ref{d0L1a}) into (\ref{firstorderinvariance}). Note that there is no room for deformations with higher powers of fluxes, as those would be of higher order in $\alpha'$. Reinserting the parameter $b$, we obtain the full first order corrections to the $\beta$-transformations in the generalized Bergshoeff-de Roo scheme 
\begin{subequations}\label{betat}
\begin{align}
\delta^{(1)} e_{a b} &= \frac {a + b} 8 \beta_{(a}{}^e \left( \omega_{b) c d} H_{e}{}^{c d} + H_{b)c d} \omega_{e}{}^{c d} \right) +  \frac {b - a} 4 \beta_{(a}{}^e \left( \omega_{b) c d} \omega_{e}{}^{c d} + \frac 1 4 H_{b)c d} H_{e}{}^{c d} \right) \ , \  \\
\delta^{(1)} b_{a b} &=    (a + b) \left[ \beta^{e c} \omega_{e [a}{}^d \omega_{b] c d} - \beta^{e c} \omega_{[\underline{a} e}{}^d \omega_{\underline{b}] c d} - \frac 1 2 \beta_{[a}{}^c \omega_{b] d e} \omega_c{}^{de} - \frac 1 8 \beta_{[a}{}^c H_{b] d e} H_{c}{}^{d e}\right] \nn\\
& + \, \frac {b - a} 2 \left[ \beta^{e c} \omega_{e [a}{}^d H_{b] c d} - \beta^{e c} \omega_{[\underline{a} e}{}^d H_{\underline{b}] c d} - \frac 1 2 \beta_{[a}{}^c \omega_{b] d e} H_c{}^{de} - \frac 1 2 \beta_{[a}{}^c H_{b] d e} \omega_{c}{}^{d e}\right] \ .
\end{align}
\end{subequations}

We have verified that these transformations preserve the action to all orders in the flux expansion. Interestingly, one can check that in fact the Lagrangian itself is invariant. As a final test we have  also verified that these transformations close in combination with the local symmetries of the theory, with respect to the following $\alpha'$-corrected brackets 
\begin{align}
\Lambda_{12}{}_{a b} &= 2 \, \beta_{1[a}{}^{c} \beta_{2\, b] c} 
    + 2\, \Lambda_{1 [a}{}^{c} \Lambda_{2\, b] c}
    + 2\, \xi_{[1}^{\mu}\partial_{\mu}\Lambda_{2] a b} \cr
  &\ \ \ -  4\,{F}_{c [a} \beta_{[1}{}_{b]}{}_{d} \beta_{2]}^{c d}  
- 4\, {F}^{c d} {{\beta_{1}}}_{c [a} {{\beta_{2}}}_{b] d}  
- \left[\frac{a+b}{4} \,  {H}_{ec d}   
+\frac{b-a}{2} \, {\omega}_{ec d}  \right]\beta^e_{[1  [a}\, {D}_{b]} \Lambda_{2]}^{c d} 
\cr
& \ \ \ - \left[(a+b)\, {\omega}^{e f}{}_{d}  + \frac{b-a}{2} {H}^{e f}{}_{d}
\right] \left({\omega}_{[a}{}^{c d}  + {\omega}^{cd}{}_{[a} \right) \beta_{[1 b] e} \beta_{2] c f} \ ,  \cr
\lambda_{12 \mu}&=
4 \,\xi_{[1}^{\nu} \partial_{[\nu}\lambda_{2] \mu]}  
- \, \frac{a+b}{2} \left(
 \Lambda_{[1}^{a b}\; \partial_{\mu} \beta_{2]}{}_{a b} 
 - 2\; \beta_{[1}^{a b} \; \partial_{\mu}\Lambda_{2]}{}_{a b} \right)  \cr 
& \ \ \ - (b-a) \left( \Lambda_{[1}^{a b} \partial_{\mu}\Lambda_{2]}{}_{a b} 
+ \beta_{[1}^{a b} \partial_{\mu}\beta_{2]}{}_{a b} 
\right) \ , \cr 
\xi_{12}^{\mu} &= 2 \xi^{\nu}_{[1} \partial_{\nu}\xi_{2]}^{\mu} + \beta_{[1}^{\mu\nu} \lambda_{2] \nu} \; ,
\end{align}
where  $\xi$ is the vector that generates diffeomorphisms  and $ \lambda$ is the one-form that generates the gauge transformations of the two-form, and we have also defined
\begin{eqnarray}
F_{ab}=\frac{a+b}8\left(\omega_{acd}\omega_b{}^{cd}+\frac14H_{acd}H_b{}^{cd}\right)+\frac{b-a}8\omega_{(a}{}^{cd}H_{b)cd}\ . 
\end{eqnarray}

\subsection*{Acknowledgements} 

We warmly thank J.J. Fernandez-Melgarejo for collaboration in the initial steps of this work. In addition we are very grateful to O. Hohm and T. Codina for comments on the manuscript.  Support by Consejo Nacional de Investigaciones Cient\'ificas y T\'ecnicas (CONICET), Agencia Nacional de Promoci\'on Científica y T\'ecnica (ANPCyT), Universidad de La Plata (UNLP) and Universidad de Buenos Aires (UBA) is also gratefully acknowledged.

\begin{appendix}
\section{Appendix}

We use  $\mu,\nu,\rho,\dots$ and  $a,b,c,\dots$ indices for space-time and tangent space coordinates, respectively. The infinitesimal Lorentz transformations of the vielbein and two-form are
\be
\delta_\Lambda e_\mu{}^a = e_\mu{}^b \Lambda_b{}^a \ \qquad {\rm and} \ \qquad \delta_\Lambda b_{ab}=-\frac{a+b}4\partial_{[a}\Lambda^{cd}H_{b]cd}+\frac{a-b}2\partial_{[a}\Lambda^{cd}w_{b]cd}\, .
\ee
The former allows to introduce a spin connection 
\be
\omega_{c a b} = \Omega_{[c a] b} - \Omega_{[c b] a} - \Omega_{[a b] c} \ , \ \ \ {\rm with} \ \ \ \Omega_{a b}{}^c = e^\mu{}_a \partial_\mu e_\nu{}^c \, e^\nu{}_b \, ,
\ee
that turns flat derivatives $D_a$ into covariant flat derivatives $\nabla_a$
\be
\nabla_a T_b = D_a T_b + \omega_{a b}{}^c T_c \ , \ \ \ D_a = e^\mu{}_a \partial_\mu\, ,
\ee
due to its Lorentz transformation
\be
\delta_\Lambda \omega_{c a b} =  D_c \Lambda_{a b} + \omega_{d a b}  \Lambda^d{}_c  + 2 \omega_{c d [b} \Lambda^d{}_{a]}\, .
\ee

All the expressions in the paper are flat index valued. For instance, the three-form is defined as
\be
H_{a b c} = 3\, e^\mu{}_a e^\nu{}_b e^\rho{}_c\, \partial_{[\mu} b_{\nu\rho]} \ ,  \ \ \ \ \nabla_{[a} H_{bcd]} = 0 \ , \label{BI1}
 \ee
and the Riemann tensor  is 
\be
R_{a b c d} = 2 D_{[a}\omega_{ b] c d} + 2 \omega_{[a b]}{}^e \omega_{e c d} + 2 \omega_{[\underline{a} c}{}^e \omega_{\underline{b}] e d}\, .
\ee
While the symmetry $R_{a b c d} = R_{[ab] [cd]}$ is manifest, other symmetries of the Riemann tensor are hidden and  determine the Bianchi identities
\be
R_{a b c d} = R_{c d a b} \ , \ \ \  \ R_{[a b c] d} = 0 \ . \label{BI2}
\ee
The Ricci tensor and scalar curvature are given by the traces
\be
R_{a b} = R^c{}_{a c b} \ , \ \ \  R = R_a{}^a \ , 
\ee
and since the symmetry  of $R_{a b}$ is not manifest, there is a new Bianchi identity
\be
R_{[a b]} = 0\ . \label{BI3}
\ee

\noindent The explicit expression of the tensor $T^a$ that appears in \eqref{d0L1a} is 
 \begin{footnotesize}
\begin{eqnarray}
T^a &=&  - \frac{1}{16}\, {D}_{b}{{H}^{b c d}}\,  {H}_{c d e} {\beta}^{a e} -
\frac{1}{16}\, {D}_{b}{{H}_{c d e}}\,  {H}^{b c d} {\beta}^{a e} +
\frac{1}{8}\, {D}_{b}{{H}^{b c d}}\,  {\beta}^{a e} {\omega}_{e c d} +
\frac{1}{8}\, {D}_{b}{{H}_{c d e}}\,  {\beta}^{a c} {\omega}^{b d e} \label{ta} \\ &&+
\frac{1}{8}\, {D}_{b}{{\omega}^{b c d}}\,  {H}_{c d e} {\beta}^{a e}  +
\frac{1}{8}\, {D}_{b}{{\omega}_{c d e}}\,  {H}^{b d e} {\beta}^{a c} -
\frac{1}{4}\, {D}_{b}{{\omega}^{b c d}}\,  {\beta}^{a e} {\omega}_{e c
d} - \frac{1}{4}\, {D}_{b}{{\omega}_{c d e}}\,  {\beta}^{a c}
{\omega}^{b d e} \nonumber \\ &&+ \frac{1}{8}\, {D}_{b}{\phi}\,  {H}^{b c d} {H}_{c d
e} {\beta}^{a e} - \frac{1}{4}\, {D}_{b}{\phi}\,  {H}^{b c d}
{\beta}^{a e} {\omega}_{e c d}  - \frac{1}{4}\, {D}_{b}{\phi}\,  {H}_{c
d e} {\beta}^{a c} {\omega}^{b d e} + \frac{1}{2}\, {D}_{b}{\phi}\,
{\beta}^{a c} {\omega}^{b d e} {\omega}_{c d e} \nonumber \\ &&- \frac{1}{16}\,
{H}^a{}_{b c} {H}^{b d e} {H}_{d e f} {\beta}^{c f}  + \frac{1}{4}\,
{H}^{a b c} {H}_{b d}{}^{ e} {\beta}^{d f} {\omega}_{c e f} + \frac{1}{4}\,
{H}^{a b c} {H}_{b d e} {\beta}^{d f} {\omega}_{f c}{}^{ e}  - \frac{1}{8}\,
{H}^a{}_{ b c} {H}_{d e f} {\beta}^{b d} {\omega}^{ce f} \nonumber \\ & & +
\frac{1}{16}\, {H}^{b c d} {H}_{b c e} {\beta}^{a e} {\omega}^f{}_{d f}%
  + \frac{1}{8}\, {H}_{b c d} {H}^{b c}{}_{ e} {\beta}^{e f} {\omega}_{f}{}^{
a d}   + \frac{1}{4}\, {H}^a{}_{ b c} {\beta}^{b d} {\omega}^{c e f}
{\omega}_{d e f} + \frac{1}{2}\, {H}^{a b c} {\beta}^{d e} {\omega}_{b
d f} {\omega}_{c e }{}^{f} \nonumber \\ && - \frac{1}{8}\, {H}_{b c d}
{\beta}^{a b} {\omega}_{e}{}^{ c d} {\omega}_{f}{}^{ e f}  + \frac{1}{8}\, {H}^{b
c d} {\beta}^{a e} {\omega}_{b e f} {\omega}^f{}_{c d}  - \frac{1}{8}\,
{H}^{b c d} {\beta}^{a e} {\omega}_{e b c} {\omega}_{f d}{}^{ f} -
\frac{1}{8}\, {H}^{b c d} {\beta}^{a e} {\omega}^f{}_{ b c} {\omega}_{f d
e}\nonumber \\ && - \frac{1}{4}\, {H}_{b c d} {\beta}^{b e} {\omega}_{e }{}^{a f}
{\omega}_{f}{}^{ c d}  + \frac{1}{4}\, {H}_{b c d} {\beta}^{e f}
{\omega}_{e}{}^{ a b} {\omega}_{f}{}^{ c d} - \frac{1}{2}\, {H}^{a b c} {\beta}^{d e}
{\omega}_{b d f} {\omega}_{e c }{}^{f} + \frac{1}{4}\,
{\beta}^{a b} {\omega}_{b c d} {\omega}_{e}{}^{ c d} {\omega}_{f}{}^{ e f} \nonumber \\ &&  -
\frac{1}{4}\, {\beta}^{a b} {\omega}^c{}_{ d e} {\omega}_{f}{}^{ d e}
{\omega}^f{}_{ b c} - \frac{1}{2}\, {\beta}_{b c} {\omega}^{b a d}
{\omega}^{c e f} {\omega}_{d e f} + \frac{1}{8}\,
{H}_{a b c} {H}^{b d e} {\beta}^{c f} {\omega}_{f d e}+   \frac{1}{16}\, {H}_{b c d} {H}^{b c e} {\beta}^{a f} {\omega}^d{}_{e
f}  +\dots \nonumber
\end{eqnarray} 
\end{footnotesize}
where the dots represent terms of quartic order in fluxes.

\end{appendix}

\end{document}